\documentclass[12pt]{article} 
\newcommand{\be}{\begin{equation}} 
\newcommand{\ee}{\end{equation}}

\newcommand{\ra}{\rightarrow} 
\newcommand{\bea}{\begin{eqnarray}} 
\newcommand{\eea}{\end{eqnarray}} 
\newcommand{\Tr}{{\rm Tr}}
\newcommand{\STr}{{\rm STr}}

\newcommand{\we}{\wedge}
\newcommand{\w}{\omega}
\newcommand{\N}{{\cal{N}}} 
 
\newcommand{\tilda}{\tilde} 
\begin{document} 

\begin{flushright}
hep-th/0111156
\end{flushright}

\bigskip \bigskip 
\centerline{\large \bf The polarization of F1 strings into D2 branes:}
\medskip 
\centerline{\large \bf ``Aut Caesar aut nihil \footnote{Latin for ``either Caesar or nothing''}''} 
\bigskip 

\bigskip       
\centerline{{\bf Iosif Bena}}  
\medskip 
\centerline{ Department of Physics } 
\centerline{University of California} 
\centerline{Santa Barbara, CA  93106}
\medskip 
\centerline{{\rm iosif@physics.ucsb.edu} }
\bigskip \bigskip 

\begin{abstract}         
We give matrix and supergravity descriptions of type IIA F-strings
polarizing into cylindrical D2 branes. When a RR four-form field strength 
$F_4$ is turned on in a supersymmetric fashion (with 4 supercharges) , a complete analysis 
of the solutions reveals the existence of a  moduli space of F1 $\ra$ D2 polarizations (Caesar) 
for some fractional strengths of the perturbation, and of no polarization whatsoever (nihil) 
for all other strengths of the perturbation. This is a very intriguing phenomenon, whose 
physical implications we can only speculate about. In the matrix description of the polarization we
use the Non-Abelian Born-Infeld action in an extreme regime, where the commutators of the fields are much
larger than 1. The validity of the results we obtain, provides a direct confirmation of 
this action, although is does not confirm or disprove the symmetrized trace prescription. 

\end {abstract} 
\newpage

\section{Introduction} 
Recently, several papers have appeared describing states where strings and branes  are polarized into higher 
dimensional branes, following the ideas in \cite{myers}. Usually the higher brane is supported 
against collapse by the presence of the field with which it couples. Nevertheless, only rarely is the 
back-reaction of this field on the other supergravity fields taken into account. This is because 
taking back-reaction into account does not considerably change the physics in the case of polarization 
into objects two or more dimensions higher. Nevertheless, when one
studies the polarization of a string or brane into a brane one dimension higher (like  F1 $\ra$ D2 or 
D4 $\ra$ NS5) the effect of this back-reaction on the physics is crucial.

The polarization of $N$ long fundamental strings into a tubular D2 brane in the presence of a 
four-form background $F_{0xyz}=m$ was 
first studied in \cite{emparan}.
Ignoring back-reaction, the potential was found to be of the form
\be
{V(R) \over L} = {1 \over 2 \pi g_s} \left(\sqrt{R^2 + g_s^2 N^2} - {m\over 2} R^2 \right),\label{vnaive}
\ee
where $R$ is the radius of the tube. This naive potential seems to imply that for any $F_{0xyz} $ the 
potential is unbounded below at large $R$, and the strings have a finite life time as they can
tunnel into a D2 brane. Furthermore, if one expands the square root for $g_sN > R$, 
\be
{V(R) \over L} \approx {1 \over 2 \pi g_s} \left( g_s N + {R^2 \over 2 g_s N} - {m \over 2} R^2   
\right).\label{vnaive2}
\ee
one can also see that for $F_{0xyz} >{1 \over  g_sN} $ this potential implies that the strings are not stable 
classically. Thus, a large number of coincident F1 strings could spontaneously disappear from the 
spectrum when a tiny $F_4$ is present. We got this puzzling result because the  back-reaction of $F_4$ on 
the dilaton and graviton was not taken into account. This back-reaction gives a term in (\ref{vnaive2}) 
of the form
\be
{V_{back-reaction}(R) \over L}  \sim  N m^2 R^2,
\ee
which usually prevents the decay of the strings. Finding the exact coefficient of this term is 
crucial if one wants to understand the physics of this system. A very large coefficient 
may prohibit polarization entirely, while a very small one leaves the puzzle about the strings 
disappearing at infinity unresolved.

There are several way to find this coefficient. The hard and straightforward way is to compute the 
back-reaction explicitly using the supergravity equations of motion. This has been done for other polarization 
cases in \cite{fm,diana}.
A slicker approach is to use supersymmetry  \cite{ps,m2,d2} and to find this coefficient by 
completing the square in the 
less naive version of (\ref{vnaive2}). One can ascertain how much supersymmetry is preserved by this $F_4$ 
background either by doing a supergravity computation \cite{mariana,mariana2} or by relating this
background to a boundary field theory via an AdS-CFT type correspondence \cite{maldacena,imsy}. 
Since the latter method only involves examining the supersymmetry of 
a two dimensional field theory, we will be using it here. Chapter 3 will be devoted to this subject.

Another aspect which makes the study of F1 $\ra$ D2 polarizations transparent is the existence of a 
matrix description for the resulting state. This is the subject of chapter 2.
As shown in \cite{myers} and expanded in \cite{ps}, Dp $\ra$ 
D(p+2) polarizations can be understood in terms of the matrices describing the position of the Dp branes 
becoming noncommutative. This description lacks however in other polarization cases.

At first glance, to give a matrix description to  F1 $\ra$ D2 polarizations involves using matrix string 
theory \cite{dvv}. \footnote{In a recent preprint a  matrix string theory description of the  D1/F1 string
polarizing into a D5 brane was atempted \cite{m}}
 Nevertheless, a much simpler way to describe these polarizations is to add a small 
number of D0 branes to the configuration, and to use the D0 brane noncommutative coordinates to describe it. 
Our system can be recovered  by taking the D0 brane density to zero \footnote{We thank W. Taylor for this 
suggestion.}. 

This configuration of strings blowing up into a tubular D2 brane in the presence of D0 brane 
charge was considered recently 
by Mateos and Townsend \cite{supertubes}. In a following paper \cite{bak}, a Matrix Model description 
for the supertubes was found. 
Nevertheless, this Matrix Model description does not work in the limit we are interested in (when the D0 
contribution to the energy vanishes). To describe this limit one must use the full Non Abelian Born Infeld 
action of the D0 branes \cite{tvr2,myers,tvr1,nbi}. We should note that in this limit the field strength
$F$ is much larger than one. 

It is interesting to ask whether the validity of the
result we get (which agrees exactly with the supergravity result and the one in \cite{supertubes}) confirms the
validity of the symmetrized trace prescription. The answer is probably negative, because our $F$'s are very 
simple, 
and their commutators vanish. Thus the computation is probably insensitive to the inability of the 
symmetrized trace
prescription to produce the right terms when commutators of F's are involved \cite{nbp}.

Anticipating, we will find that the polarization of F1 strings into a tubular brane is impossible except for 
values of $F_4$ proportional to fractions whose denominator is smaller than $N$. For these values 
a number of strings polarizes into one or more tubular D2 branes. The radius of these polarization configuration
is a modulus.

\section{Matrix description}

It is possible to describe F strings with D0 charge blown up into a tubular D2 brane using the D0 brane 
matrix degrees of freedom. This description is valid  both with and without an $F_4$ flux turned on 
\cite{unpublished}.
In order to give an ensemble of D0 branes a D2 charge, one has to take an ansatz where three of the D0 
brane coordinates are not commuting.  F1 charge can be given as well, by having
the noncommuting coordinates be time dependent. This is expected both from the form of 
the nonabelian coupling with $B_2$, which contains a time derivative of the coordinates, and also by noticing 
\cite{supertubes} that a tubular D2 brane with string and D0 charge has a nonzero angular momentum.

Let us assume to have $N$ D0 branes uniformly distributed along the $z$ direction. An ansatz for 
the $N \times N$ matrices which gives the right charges is \cite{unpublished}
\bea
Z_{ij} = l \delta_{ij} j ,\ \ \  X = R (a+a^{\dagger}), \ \ \ \ Y = iR (a^{\dagger}-a), \nonumber \\
a_{ij} = e^{-i\w t} \delta_{i-1,j}, \ \ \ a_{ij}^{\dagger} = e^{i\w t} \delta_{i+1,j}.  
\eea
In the infinite volume limit, the matrices $X,Y$ and $Z$ are infinite, and obey several useful identities:
\bea
-i F_{zx}\equiv [Z,X]=-ilY, \ & &- i F_{zy}\equiv [Z,Y]=ilX, \ \ \-i F_{xy}\equiv [X,Y]= 0 \nonumber \\
F_{tx}\equiv \dot{X} = \w Y , & & \   F_{ty}\equiv \dot{Y} = -\w X.
\label{f}
\eea
Note that we work in the normalization where the matrices are dimensionless. To obtain the physical quantities corresponding to dimensionless symbols (such as $R, l$ and $\w$) one has to multiply by the appropriate powers of $l_s = \sqrt{2 \pi \alpha'} $. 
Since $l$ is the separation between the individual D0 branes, $lN$ is the size of the $z$ direction, which 
we take to be very large. The local D0 charge density is $Q_0 = 1/l$. For weak fields the F string charge density along the $z$ direction is given by \cite{tvr1}
\be
Q_1= {T_0 l_s  \over i l N} \STr \left( \dot{X}^i [X^i, Z] \right) = 4 T_0 l_s \w  R^2, 
\label{q1}
\ee
and is quantized.  As we will see shortly, this quantization proceeds differently for 
large string charge.

The D2 dipole charge is proportional to 
\be
{1 \over - il N} \Tr \left( X [Y, Z] + Y [Z,X]+ Z [X,Y])\right) =  4 R^2. 
\ee
The first two terms in the left hand side are nonzero, while the third is zero. This means there is local D2 charge in the $xz$ and $yz$ planes, but not in the $xy$ plane, exactly as it should be for a cylindrical D2 brane.

As the matrices $X,Y$, and $Z$ are infinite in the large volume limit, the ciclicity of the trace is not respected.
Without any $F_4$ turned on, the action is given \cite{tvr2} by 
\be
L_{NBI} = - T_0 \ {\rm STr} \left(\sqrt{-{\rm det} \left( \begin{array}{cccc}
-1 & F_{tx} & F_{ty}& 0 \\
-F_{tx}&1&0&F_{xz} \\
-F_{ty}&0&1&F_{yz}\\
0&-F_{xz}&-F_{yz}&1
 \end{array}\right)} \right).
\ee
One can plug in the $F_{ij}$ defined in (\ref{f}), and evaluate the determinant above. By using the fact that the determinant of a matrix with noncommuting entries \footnote{The determinant is defined as ${\rm det}\ M = \epsilon^{a_1a_2 ... a_n} \epsilon^{a'_1a'_2 ... a'_n} M_{a_1a'_1}M_{a_2a'_2} ...  M_{a_na'_n}$} takes all orderings into account, or alternatively the fact that $X$ and $Y$ commute, we notice that the terms proportional to $X^2 Y^2$ and their permutations cancel. The only combination of matrices which appears under the square root is $X^2+Y^2$. Since this combination is proportional to $ {\bf 1}$, the Lagrangian simply becomes
\be
L= - T_0 \ N \sqrt{1 + 4 l^2 R^2 - 4 \w^2 R^2}.
\ee
In order to obtain a Hamiltonian from this Lagrangian, one has to find the conjugate momentum corresponding to the string charge:
\be
Q_1 = {l_s \over  N} {\delta L \over  \delta \w} = {4 T_0 l_s  \w R^2 \over \sqrt{1 + 4 l^2 R^2 - 4 \w^2 R^2}},
\label{Q1}
\ee
where the normalization of the string charge with respect to ${\delta L \over  \delta \w} $ was found by comparing the small string charge limit of (\ref{Q1}) with (\ref{q1}). The Hamiltonian is therefore
\be
H = \w {\delta L \over  \delta \w} - L = T_0 N \sqrt{1+ {Q_1^2 g_s^2 \over 8 \pi R^2}} \sqrt{1+4 l^2 R^2},
\ee
the same as the one found in \cite{supertubes}. This formula is valid for all values of the D0, F1 and D2 charges. We can take the limit when the D0 charge vanishes ($N \ra 0$ keeping the length of the tube $l N$ fixed) and the string charge dominates to obtain
\be
{H \over N l l_s} = {Q_1 \over l_s^2} + {4 \pi R^2 \over l_s Q_1 g_s^2}. \label{H}
\ee
The first term is the energy of $Q_1$ coincident static strings, and the second one reproduces exactly the Born-Infeld contribution to the action (\ref{bi}) when one identifies the parameters in the matrix and supergravity descriptions:
\be
r^2 = x^2 + y^2 = {\STr (X^2+Y^2) \over N} l_s^2 = 4 R^2 l_s^2.
\ee
This exact agreement provides a confirmation of the nonabelian Born-Infeld action, as well as of the 
nonabelian coupling with the $B$ field found in \cite{tvr1}. Nevertheless, it does not imply anything about the 
validity of the symmetrized trace prescription because the only matrix whose trace we had to take was $ {\bf 1}$.

The presence of $F_4$ modifies the action by the addition of
\be
L_{F_4} = T_0 \STr(C_{0zx}F_{zx} +C_{0zy}F_{zy} )= 
 F_{0xyz} 2 l N T_0 R^2. \label{FL} 
\ee
The terms in (\ref{H}) and (\ref{FL}) correspond to the terms found in the naive action (\ref{vnaive2}) .
Unfortunately, the back-reaction term cannot be found in the matrix description because 
the $(F_4)^2$ couplings in the nonabelian D brane action are not known.

\section{The supergravity analysis}

Let us for the beginning explore the supergravity background created by a large number of coincident 
F strings. By a 
generalization of  the AdS-CFT correspondence in the spirit of \cite{imsy}, the two-dimensional field 
theory  living on the worldvolume of these coincident F strings is dual to string theory living in the near-horizon background of these strings. As explained in \cite{bdhm-bklt}, turning on an operator in the 
boundary theory Hamiltonian corresponds to turning on a nonnormalizable mode in the bulk. One can 
study the effect of the operator on the boundary theory by examining the supergravity dual which the 
corresponding nonnormalizable mode creates. Nevertheless the study of this theory, although 
straightforward once all the computations are done, is not our primary interest. Rather, we are looking for 
a supergravity background with a nonzero $F_4$ which preserves 4 of the original 16 supercharges, and 
we are using this correspondence only as a tool to keep control on the supersymmetry.

As the fermions of the IIA F string in static gauge transform in the  {\bf8s} of the SO(8) R symmetry group, a 
fermion 
bilinear transforms in the {\bf35}$_+$ or the {\bf 28} of this group. The bilinear in the {\bf 35}$_+$ 
corresponds to a self 
dual RR 4-form field strength on the space transverse to the strings. This 4-form polarizes the strings into 
D4 branes, creating the dimensional reduction of the M2$\ra $M5 setup in \cite{m2}.
The fermion bilinear in the {\bf 28} corresponds to a bulk RR 2 form and 6 form field strengths on the 
transverse space. 
The 6 form on the transverse space is equivalent by Hodge duality to a 4 form with 2 legs along the 
strings, which is what we need for F1$\ra$ D2 polarization.

{\bf The background}

The string frame supergravity background of $N$ fundamental strings is
\bea
ds^2& =& Z^{-1}(-dx_0^2+dx_1^2) + dx_{\perp}^2 \nonumber \\
e^{\Phi}&=& g_s Z^{-1/2} \nonumber \\
B &=& Z^{-1} dx^0 \wedge dx^1, \label{f1}
\eea
where $Z$ is a harmonic function on the transverse space which becomes 
\be
Z = 1+ {R^6 \over r^6},\ \ \ R^6 = 32 \pi^2 N g_s^2 \alpha'^3, 
\ee
when the strings are coincident. If $g_s > 1$, the supergravity solution is everywhere valid and it is not dual 
to anything. However, for the beginning we are interested in a regime where we could use the AdS-CFT type of duality discussed above, and thus we will focus for now on $g_s > 1$ \footnote{Once we establish the form of the supersymmetric supergravity perturbation, we can also use it for $g_s < 1$ (as long as the 1 in the harmonic function can be ignored) since the supersymmetry of the background does not depend on $g_s$.}.
For $r^6 > g_s^2 N$ the dilaton becomes large, and physics starts being described by the supergravity solution of N coincident M2 branes. For $1 <  r^6 < g_s^2 N$ supergravity in the background (\ref{f1}) gives a weakly coupled description of the physics, and the 1 in the harmonic function can be ignored. We will work in this regime. For even smaller radii, $r<1$ the theory has a weakly coupled description in terms of an orbifold CFT, analyzed in \cite{dvv}.

{\bf The perturbation}

We are interested in perturbing the above background with the RR fields which could cause 
polarization. As explained above, these fields are $F_2$ and $F_6$ on the space transverse to the strings.
If one defines \hspace{-.2cm} \footnote{This is not the canonical definition of 
$F_6$ but it is the most convenient to work with.} $\ \ F_6 \equiv *(F_4-C_1 \we H_3)$, the IIA equations of motion become:
\bea
d*F_2-H_3 \we F_6 =0 \nonumber \\
dF_6 - H_3 \we F_4 =0 \nonumber \\
- d * F_6 = d (F_4-C_1 \we H_3) = -F_2 \we H_3 \nonumber \\
d F_2 =0 \label{eom}
\eea
Combining (\ref{f1}) with  (\ref{eom}) and expressing the ten dimensional Hodge dual $*$ in terms of 
the 8 dimensional transverse space Hodge dual $*_8$, the equations satisfied by the first order 
perturbations become:
\bea
d\left[Z^{-1}(*_8 F_2 - F_6)\right] = 0, \hspace{1in} 
d F_6 = 0, \nonumber \\
d\left[Z^{-1}(*_8 F_6 - F_2)\right] = 0, \hspace{1in} 
d F_2 = 0.
\label{eom2}
\eea
We should note that acting on even forms $*_8^2=1$, while $*^2=-1$.
To find the perturbations with the same R-symmetry transformation properties as the fermion bilinear 
it is convenient to group the 8 transverse coordinates and the fermions into 4 complex pairs:
\bea
z^1 = x^2 + i x^3\ ,\ \  z^2 = x^4 + i x^5\ ,& &  z^3 = x^6 + i x^7 ,\ \ z^4 = x^8 + i x^9, \nonumber \\
\Lambda^1 = \lambda^1 + i \lambda ^2\ ,\ \  \Lambda^2 = \lambda^3 + i \lambda^4 ,& & \Lambda^3 =\lambda^5 + i\lambda^6\ ,\ \ \Lambda^4 = \lambda^7 + i \lambda^8 \ .
\eea
Under the  rotations $z^i \rightarrow e^{i \phi_i} z^i$ the fermions transform as :
\bea
\Lambda^1 & \rightarrow & e^{i(\phi_1-\phi_2+\phi_3+\phi_4)/2 } \Lambda^1  \nonumber \\
\Lambda^2 & \rightarrow & e^{i(\phi_1+\phi_2+\phi_3-\phi_4)/2 } \Lambda^2 \nonumber \\
\Lambda^3 & \rightarrow & e^{i(\phi_1-\phi_2-\phi_3-\phi_4)/2 } \Lambda^3 \nonumber \\
\Lambda^4 & \rightarrow & e^{i(\phi_1+\phi_2-\phi_3+\phi_4)/2 } \Lambda^4. 
\eea
Thus a fermion bilinear of the form ${\rm Re} (\bar \Lambda^1 \Lambda^2)$ has the same SO(8) transformation
properties as
$T_2 = {\rm Re} (d z ^2 d \bar z^4)$, $V_2$ (defined in the Appendix) and their Hodge duals $T_6$ and $V_6$ \footnote{The standard properties  of these tensors are reviewed in the Appendix.}. 

The most straightforward way to examine the supersymmetry preserved by a specific fermion bilinear is to examine 
the free 2-dimensional boundary theory which has 8 bosons $X_{\mu}$ in ${\bf 8_v}$ of SO(8) and both the left and right moving fermions $\lambda_a$ and $\tilde\lambda_a$ in ${\bf 8_s}$. The original supersymmetry transformations are
\bea
\delta X^{\mu} &=& -\epsilon_{\dot a} \Gamma_{\dot a a}^{\mu} \lambda_a 
- \tilda \epsilon_{\dot a} \Gamma_{\dot a a}^{\mu} \tilda \lambda_a  \nonumber \\
\delta \lambda_a &=& \partial X_{\mu} \Gamma_{a \dot a }^{\mu}\epsilon_{\dot a} 
+  F_{\mu} \Gamma_{a \dot a }^{\mu} \tilda \epsilon_{\dot a} \nonumber \\
\delta \tilda \lambda_a &=& \bar \partial X_{\mu} \Gamma_{a \dot a }^{\mu} \tilda \epsilon_{\dot a} 
-  F_{\mu} \Gamma_{a \dot a }^{\mu}\epsilon_{\dot a}\nonumber \\
\delta F^{\mu} &=& -  \tilda \epsilon_{\dot a} \Gamma_{\dot a a}^{\mu} \bar \partial \lambda_a 
+ \epsilon_{\dot a} \Gamma_{\dot a a}^{\mu} \partial \tilda \lambda_a,
\eea
where $\Gamma_{\dot a a}^{\mu} $ are $8 \times 8$ matrices given in \cite{gsw}. It is a straightforward 
exercise to show that adding a fermion bilinear, ${\rm Re} (\bar \Lambda^1 \Lambda^2)$ together with its 
bosonic partner preserves 4 of the original 16 supercharges $\epsilon_{\dot a}$ and $\tilda \epsilon_{\dot a}$.
\footnote{On can also examine the supersymmetry preserved by a certain fermion bilinear by doing a  9-11 
flip. A bilinear of the form $\bar \Lambda ^1 \Lambda^2$ becomes a vector bilinear on the M2 brane when lifting to 
M-theory. Reducing along a different direction, this becomes a vector bilinear on the D2 brane, of the 
form $\bar \lambda^1 \gamma^{\mu} \lambda^2$ which upon T duality along $\mu$ becomes again a scalar fermion 
bilinear on the D1 brane: $\bar \lambda^1 \gamma^{5} \lambda^2$. Reexpressing the complex fermions in real ones, this bilinear is $\lambda^1 \lambda ^3 - \lambda^2 \lambda^4$, which can be expressed as coming from a superpotential of the form $W = \phi^1\phi^3 - \phi^2 \phi^4$. Thus after changing the sign of a field this superpotential becomes the dimensional reduction of a hyper bilinear in the 2D SYM on the D2 branes. Since that superpotential preserves 4 supercharges ($\N=2$ in 2+1 dimensions), this is the supersymmetry preserved by our operator.} 

Thus, a 2 form and a dual 6 form corresponding to the fermion bilinear can be written as combinations of $T_2$ 
and $V_2$, and respectively $T_6$ and $V_6$ multiplied by powers of $r$ . Plugging this ansatz in (\ref{eom2}), we find 4 solutions. Two of them correspond to a mode which is not a chiral primary, and the other two, one normalizable and the other non normalizable correspond to the fermion bilinear we are interested in. Adding the fermion bilinear to the boundary theory Hamiltonian corresponds  \cite{bdhm-bklt} to adding to the background (\ref{f1}) the non normalizable solution:
\be
F_2 = {m \over \sqrt{\alpha'}}   Z (2 T_2 -6 V_2),  \ \ \ \  
F_6 = - {m \over \sqrt{\alpha'}} Z (6 T_6 - 6 V_6),
\ee
where $m$ is proportional to the coefficient of the fermion bilinear added to the boundary Lagrangian.

A tubular D2 brane in this geometry couples with $C_3 + B_2 \we C_1$. In our background, this 
combination can be found using 
\be
d (C_3 +C_1 \we B_2) = F_4 - C_1 \we H_3 + B_2 \we F_2 =  - Z^{-1} (*_8 F_6 - F_2) \we dx^0 \we dx^1 = 
2 m T_2  \we dx^0 \we dx^1 .
\label{c3}
\ee

Since   $ Z^{-1} (*_8 F_6 - F_2) $ is a harmonic form, it only depends on its value at infinity, which is given by the boundary theory. Therefore the expression above remains true when the F1 strings are coincident or when they are distributed. This 
implies that the WZ term in the action of a cylindrical D2 brane is independent on where the F1 strings 
are.

{\bf Probing the perturbed solution}

In order to find the self interaction potential of a cylindrical D2 brane with large F1 charge $N$, one has to 
find first the potential felt by a D2 brane probe with smaller F1 charge $1 \ll n \ll N$, and then to build 
the D2 
brane by bringing test D2 branes from $\infty$. Anticipating, we will find that similarly to other 
Polchinski-Strassler type problems \cite{ps,m2,d2,exotic,m5}, the probe potential will be independent of the 
distribution of the strings; thus the self interaction potential of a tubular D2 brane containing all the $N$ 
strings will be the same as the probe brane potential with $n$ replaced by $N$.

A cylindrical D2 brane can be given F1 charge by turning on a nontrivial $F_{01}$ on its worldvolume.
The quantized F1 string charge in a nontrivial $G_{\mu\nu}$, $\Phi$ and $B_{\mu\nu}$ is 
\be
\Pi = {\delta L \over \delta F_{01}} =  {2 \pi r \over \sqrt{2 \pi}(2 \pi \alpha')^{3/2} g_s} Z^{1/2} 
{2 \pi \alpha' (B_{01}+2 \pi \alpha' F_{01}) \over \sqrt{|G_{00}| G_{11} -
(B_{01}+ 2 \pi \alpha' F_{01})^2}} = n 
\ee
If we call 
\be
A \equiv {B_{01}+2 \pi \alpha' F_{01} \over \sqrt{|G_{00}| G_{11} - (B_{01}+2 \pi \alpha' F_{01})^2}}  = 
{\sqrt{2 \pi} (2 \pi \alpha')^{1/2}  n g_s \over 2 \pi r Z^{1/2} },
\ee
 the Hamiltonian density per unit string length will be
\be
H_{BI} = {2 \pi r Z^{1/2} Z^{-1} \over   \sqrt{2 \pi} (2 \pi \alpha')^{3/2} g_s  } \left( \sqrt{A^2+1} - A \right).
\ee
We are interested in probes with dominant string charge, $A \gg 1$, which is easily realized for $n > \sqrt{N}$ in the regime where supergravity is valid. In this regime the square root can be expanded, and 
the main contributions to the Hamiltonian cancel,  as they represent the interaction energy between 
parallel F1 strings. After this cancellation, the subleading term of the Born-Infeld 
action is
 \be
V_{BI}= {\pi  r^2 \over (2 \pi \alpha')^{2} n g_s^2}.
\label{bi}
\ee 
As we mentioned in the previous chapter, this supergravity result reproduces exactly the one obtained using the nonabelian Born Infeld action in the limit of vanishing D0 charge. 
For a D2 brane of $R^2 \times S^1$ geometry, this contribution to the action does neither depend on the orientation of the $S^1$, nor on $Z$. It is straightforward to generalize this action for the circle replaced by an ellipse. This can be easily done in a similar way to \cite{ps,m2,d2,exotic,m5}. If one turns on the bilinear ${\rm Re} (\bar \Lambda^1 \Lambda^2)$ which corresponds to $T_2 = {\rm Re} (d z ^2 d \bar z^4)$, it is natural to expect the probe to have a minimum if the two semiaxes of the ellipse lie in the two planes spanned by components of $ z^2 =x^4+i x^5 $ and $ z^4= x^8+ix^9$. If the lengths and orientations of the semiaxes are given by the modulus and phase of $Z_4$ and $Z_2$, the result is obtained by just replacing $r^2$ by $|Z_4|^2/2+|Z_2|^2/2$ in (\ref{bi}).

One can also find the interaction potential coming from the Wess Zumino action. Using (\ref{c3}) and integrating over the ellipsoid one finds after a few straightforward steps
\be
V_{WZ} = - {2 \pi \over (2 \pi \alpha')^{2}} {m } {\rm Re}(Z_4 \bar Z_2),
\ee 
which is also independent of the harmonic function  $Z$. Thus, the first two terms of the action of the probe are
\be
V_{BI}+ V_{WZ}  = {\pi \over 2 (2 \pi \alpha')^{2}}\left[{|Z_4|^2 + |Z_2|^2 \over  n g_s^2} - {4 m}\ {\rm Re}(Z_4 \bar Z_2) \right].
\label{v1}
\ee
As we said in the Introduction, besides these two terms there is another one coming from the backreaction of the $F_2$ and $F_6$ perturbations on the metric, dilaton and 2-form. One can see from the equations of motion that this term should be proportional to 
$m^2 g_s^2 n r^2$ 
\footnote{The factor $ m^2 g_s^2 $ comes from the square of the first order perturbation, $n$ comes from the dominant F-string charge of the probe, and the proportionality with $r^2$ is not hard to derive.}. 
Supersymmetry allows us to find this term exactly  by simply completing the square in (\ref{v1}). Thus
\bea
V_{total} &=& {\pi \over2 (2 \pi \alpha')^{2}}\left[{|Z_4|^2 + |Z_2|^2 \over  n g_s^2} 
- {4 m}\ {\rm Re}(Z_4 \bar Z_2)  +m^2 n g_s^2  \left( {|Z_4|^2 + |Z_2|^2}\right) \right] \nonumber \\
&=& {\pi \over 2 (2 \pi \alpha')^{2} n g_s^2} \left[ |Z_4 - Z_2 \ m 
n g_s^2|^2 + |Z_2 -  Z_4 \ m n g_s^2 |^2
\right].
\label{v}
\eea
 
{\bf The full potential}

The last term of (\ref{v}) is also independent of $Z$. Therefore, the potential of the probe 
tube does not depend on the position of the F1 strings sourcing the geometry. One can now find the self 
interaction potential of a D2 brane tube containing {\it all} F1 strings, in its own geometry, 
by bringing from infinity small probes and using them to construct the final configuration. Since the 
potential of each of these small probes is independent of where the rest of the probes are, the final potential 
will be the sum of the probe potentials, which gives the same result as  (\ref{v}) with $n$ replaced by $N$.
The most general configuration will consist of several D2 tubes with F1 charges $n_i$. The potential will be the 
sum of terms, each of them equal to  (\ref{v}) with  $n$ replaced by $n_i$:
\be
V_{many \ tubes} = \sum_{i}{ {\pi \over 2 (2 \pi \alpha')^{2} n_i g_s^2} \left[ |Z^i_4 - Z^i_2 \ m 
n_i g_s^2|^2 + |Z^i_2 -  Z^i_4 \ m n_i g_s^2 |^2
\right]},
\label{vf}
\ee
where $Z^i$ parametrize the semiaxes of the $i$'th D2 tube, which contains $n_i$ F1 strings.
This potential is very interesting. It implies that for a generic strength of the perturbation, there is only one minimum at $Z_4=Z_2=0$. Contrary to what one might expect, increasing the strength of the four-form does not make the polarization to a D2 brane more likely. This is because a large  $F_4$ causes a large backreaction on the graviton, dilaton and 2 form, which do not favor the polarization.

Even more remarkable, for the particular values $m = \pm {1 \over n_i g_s^2}$, the zero energy solution is given by $Z^i_4= \pm Z^i_2$ for the $i$'th tube, and $Z_4=Z_2=0$ for the others. Thus, for these values of the perturbation a moduli space of possible polarization vacua opens up; $n_i$ of the original $N$ strings can be polarized. The lowest absolute value of $m$ which allows the polarization to occur is $m = {1 \over N g_s^2}$. In this case all $N$ strings polarize. For $m = {1 \over p g_s^2}, N/2 < p < N$, $p$ of the $N$ strings polarize into one D2 brane. For $p=N/2$ it suddenly becomes possible for half the strings to polarize into one D2 brane, and the other half to polarize into another D2 brane. Both D2 tubes can be at any radius. In general, for ${N \over k-1} <p <{N \over k}$, the ground state will consist of  $k$ concentric D2 branes at different radii, with F1 charge $p$ each.

{\bf Consistency checks}

Before we continue, we should address two small but potentially dangerous issues. First, we have to make sure that the supergravity perturbation fields are weaker than the background fields. The energy of the background and perturbation fields are respectively
\bea
e^{-2 \phi} H^{(3)}_{01r} H^{(3)}_{01r} G^{00}G^{11}G^{rr} &\sim& {N \alpha'^3\over r^8}, \nonumber \\
F^{(2)}_{ij}F^{(2)}_{ij} G^{ii}G^{jj} \sim m^2 Z^2 &\sim&{m^2 N^2 g_s^4 \alpha'^5\over r^{12}}. 
\eea
The condition for validity of the perturbation expansion is thus $r^4 \gg m^2 N g_s^4 \alpha'^2$. For the smallest value of $m$ which allows for a moduli space this is equivalent to $r^4 \gg \alpha'^2/N$ which is trivially satisfied. However, for larger values of $m$ which allow $n$ strings to polarize into a D2 brane, this condition is satisfied only for  $r^4 \gg \alpha'^2 N/n^2$. 

The second issue has to do with the attraction between the opposite sides of a tube, or between concentric tubes. Because the large F1-string tension and the D2 brane tension add in quadratures, the interaction coming from the gravitational D2-D2 attraction has a negligible effect on the energy. What could spoil the interaction is the electric attraction between the opposite sides of the tube. 

 The issue of this possible interaction is usually not addressed in most of the polarization papers. In principle this can give an energy contribution which may affect the polarization ground state. In our context, since the balance of the energies is so delicate, such a contribution could in principle easily lift the moduli space. One could probably argue that supersymmetry prevents this contribution from affecting the potential, but nevertheless it is instructive to see how it happens.

As we can see from (\ref{c3}), our D2 brane tube  couples with a field combination which is proportional to
the harmonic form $ Z^{-1} (*_8 F_6 - F_2) $. The presence of extra D2 brane charge affects the Bianchi identity for $F_6$ only. Nevertheless,  $ Z^{-1} (*_8 F_6 - F_2) $ remains harmonic. Therefore the presence of D2 charge (from the opposite side of the tube) does not affect the Wess-Zumino term of the potential.
Thus, as we expected, the attraction between the opposite sides of the tube or between different tubes does not have any effect on the energy of the system. The lack of interaction  between tubes placed at different radii is characteristic to Polchinski - Strassler type setups, and is also present for the supertubes with D0 charge \cite{supertubes2}.

Usually the appearance of a moduli space of vacua signals some symmetry enhancement.
It is interesting to ask what this symmetry might be. One certain thing is that supersymmetry is not enhanced, 
since both the field strengths which allow polarization and those who do not allow it preserve 4 supercharges. 
Thus, this is probably a symmetry which appears in the UV of the mysterious orbifold CFT living on the strings.
   
\section{Conclusions and future directions}

We have given matrix and supergravity descriptions to the polarization of type IIA fundamental strings in tubular D2 branes. For the matrix description we have used the matrix degrees of freedom of D0 branes and their nonabelian Born-Infeld action to describe an F1 $\ra$ D2 tubular configuration in a nontrivial $F_4$ background.

In order to give a supergravity description while maintaining control over the supersymmetry we have turned on a fermion bilinear in the theory living on $N$ coincident fundamental strings. We have then related this bilinear to a supergravity nonnormalizable mode via an AdS-CFT like correspondence, and showed that the nonnormalizable mode induces F1 $\ra$ D2 polarization.
We have found that polarization occurs only when the value of the bulk fields is proportional to fractions 
whose denominator is smaller than $N$; morover, in these cases the polarization radius is a modulus. For any other values of the bulk fields there is no polarization.

For the supergravity description we have used the setup and technology developed by Polchinski and Strassler \cite{ps}. Although this technology is sometimes heavy, it has certain advantages  (such as the control on the supersymmetry and on the backreaction of the fields, or the lack of attraction between the opposite sides of the polarized object) which in our opinion make it the only correct setup for the study of polarization into objects one dimension higher.

One interesting question which one might ask is what happens if one turns on a nonsuper-symmetric 
traceless scalar bilinear in the potential, which might create an instablity 
along some direction. Understanding the effect of this other piece in the potential can be easily done in our framework (see \cite{d2} or \cite{k} for an example). Such a piece makes the strings cease to be a classical 
ground state, and seems to leave the puzzle in \cite{emparan} unsolved. The only consolation is that only a very special combination of 4-form and graviton will create this instability. In general however, turning on an
 $F^4$ will not afect the stability of the strings.

It is intriguing to see what is the fate of this configurations when one lifts them to M theory. A D2 brane with large F string charge lifts to an M2 brane helix which winds around the eleventh dimension many times while circulating once around an  $S^1$. The radius of the large circle is a modulus, so there is no potential for the helix to be stretched in a radially symmetric fashion. Even more puzzling to our intuition is the fact that a state of
many such helices, non concentric and possibly intertwined preserves 4 supercharges. When one takes the size of
the 11'th dimension to infinity these helices should naturally dissolve into the ``Coulomb branch'' of the 
theory on parallel M2 branes.

The F1 $\ra$ D2 configuration falls in the general category of objects polarizing into objects one dimension higher. Similar situations can occur in the case of D4 branes polarizing to NS5 branes. We expect a similar potential, which allows a moduli space to open up for a very specific value of the perturbation, and does not allow polarization for any other value. On general grounds one expects the D4 $\ra$ NS5 polarization to happen for $\N=1$ supersymmetry in 4+1 dimensions, and therefore the moduli space not to be lifted up by quantum corrections. Again, the lift of the configuration in M-theory would give a spiral-shaped M5 brane. 

We have studied a very intriguing system, and found many phenomena which challenge our intuition, such as 
the presence of a moduli space of polarization vacua or the fact that M2 brane spherical helices in M theory 
at finite $R_{11}$ can be supersymmetric. We cannot but hope that in time a more fundamental explanation of these
phenomena and other less understood aspects of M theory will be given.

{\bf Acknowledgments}: I would like to thank Wati Taylor for many useful comments and help 
in understanding the matrix description of this polarization. I would like to thank Joe Polchinski, K. S. Narain,
S. Sinha, B. de Wit, R. Roiban and A. Buchel for useful discussions. This work was supported by a
University of California Special Regents Fellowship, and by  NSF grant PHY97-22022.

\section{Appendix}

We list several properties of the transverse space antisymmetric 2- and 6-tensors 
which form a basis for the forms with the R-symmetry transformation properties of a 
fermion bilinear.

\bea
T_6& =& *_8 T_2 \\
V_6=\frac{1}{6!}(\frac{x^t x^m}{r^2} T_{tnpqrs}&+&{\rm{5\; more}})
dx^m\we dx^n \we dx^p \we dx^q\we dx^r \we dx^s, \\
V_2&=&\frac{1}{2!}(\frac{x^qx^i}{r^2}T_{qj}+{\rm{1 \;more}})
dx^i\we dx^j,\\
T_2-V_2&=&*_8 V_6, \ \ \ \ T_6-V_6 =*_8 V_2, \\
d(\ln r)\we V_6&=&0, \ \ \ \ \ d(\ln r)\we V_2=0,  \\
d(r^p(6T_6+ pV_6))&=&0,\ \ \ \ \ d(r^p(2T_2+ pV_2))=0.
\eea

\end{document}